\documentclass[conference]{IEEEtran} 

\usepackage[T1]{fontenc} 
\usepackage{amsmath,amssymb,physics} 
\usepackage{float}
\usepackage{graphicx} 
\usepackage{cite} 
\usepackage{balance} 
\usepackage{caption}
\usepackage{algorithm}
\usepackage{algpseudocode}
\usepackage{multirow}
\usepackage{booktabs}
\usepackage{url}
\usepackage{xurl}
\usepackage[dvipsnames]{xcolor}
\usepackage{orcidlink}
\usepackage{hyperref}
\definecolor{ionqorange}{HTML}{FF5000}
\hypersetup{
    colorlinks=true,
    linkcolor=ionqorange,    
    citecolor=ionqorange,    
    urlcolor=ionqorange      
}
\usepackage[capitalise]{cleveref}

\IEEEoverridecommandlockouts 
\title{Hybrid Quantum-Classical Optimization Workflows for the Shipment Selection Problem}
\author{
    \IEEEauthorblockN{
        Miguel Angel Lopez-Ruiz\IEEEauthorrefmark{1},  
        Daiwei Zhu\IEEEauthorrefmark{1},
        Jonas Hatzenb\"uhler\IEEEauthorrefmark{2},
        Shudian Zhao\IEEEauthorrefmark{2},
    }
    \IEEEauthorblockN{
        Claudio Girotto\IEEEauthorrefmark{1},
        Willie Aboumrad\IEEEauthorrefmark{1},
        Jonas Alm\IEEEauthorrefmark{2},
        Julia Kompalla\IEEEauthorrefmark{1},
        Mena Issler\IEEEauthorrefmark{1},
    }
    \IEEEauthorblockN{
        Ananth Kaushik\IEEEauthorrefmark{1},
        Martin Roetteler\IEEEauthorrefmark{1}
        \medskip
    }
    \IEEEauthorblockA{\IEEEauthorrefmark{1}IonQ Inc., 4505 Campus Dr, College Park, MD 20740, USA}
    \IEEEauthorblockA{\IEEEauthorrefmark{2}Einride AB, Stadsg\aa rden 6, Stockholm, 116 45, Sweden}
}

\begin{document}

\maketitle

\begin{abstract}
We present a quantum optimization framework for the Shipment Selection Problem (SSP) in electric freight logistics, developed jointly by IonQ and Einride. Idle gaps arising from stochastic shipment cancellations reduce fleet utilization and revenue; filling them optimally requires solving a combinatorial assignment problem with quadratic inter-gap dependencies. We formulate the SSP as a Mixed-Integer Quadratic Program, map it to an Ising cost Hamiltonian, and solve it using Iterative-QAOA, a non-variational warm-start extension of the Quantum Approximate Optimization Algorithm (QAOA) with a fixed linear-ramp parameter schedule. An end-to-end hybrid workflow integrates Einride's vehicle routing problem (VRP) solver with IonQ's quantum simulations, enabling evaluation on real, anonymized logistics data spanning up to 130 qubits. We assess solution quality through application-level performance metrics, including Shipments Delivered (SD), Schedule Compatibility Score (SCS), and Total Drive Distance (TDD). When the quantum assignment is passed to the classical solver as a warm start, the resulting hybrid workflow achieves improvements of up to 12\% in SD and a reduction of up to 6\% in total drive distance per shipment for specific instances, while total operational cost remains effectively unchanged. For the subset of instances within reach of current devices (20--35 qubits), the workflow is additionally executed on IonQ trapped-ion quantum hardware, where the hardware results closely reproduce the noiseless simulations. These results show that Iterative-QAOA can generate compatibility-aware assignments that become operationally valuable when embedded in a hybrid logistics optimization workflow.
\end{abstract}


\section{Introduction}

\begin{figure}[ht!]
    \centering
    \includegraphics[width=\columnwidth]{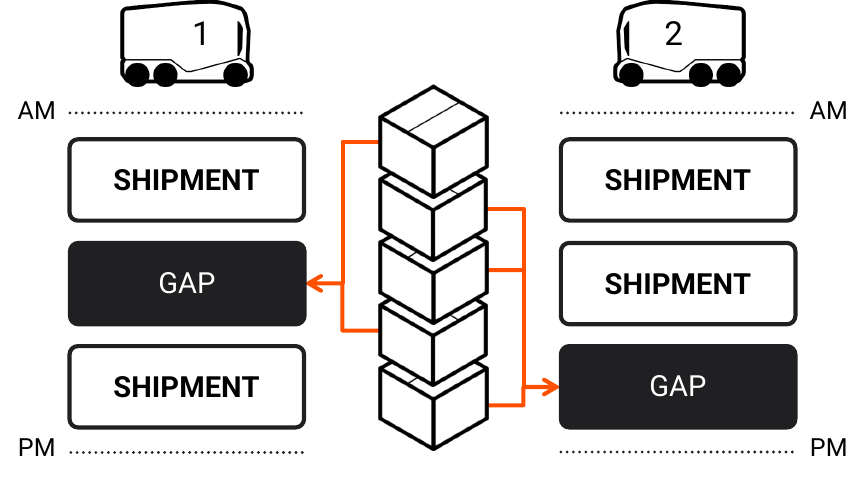}
    \caption{Illustration of the classical gap-filling scenario. Idle gaps (black blocks) in a pre-optimized truck schedule are caused by shipment cancellations. Currently, these gaps are filled from a pool of unassigned shipments (center) using Einride's state-of-the-art Electric Vehicle Routing Problem (E-VRP) solver. In this work, we present a hybrid quantum-classical workflow optimizing this step.}
    \label{fig:gap_filling_illustration}
\end{figure}

\begin{figure}[t!]
    \centering
    \includegraphics[width=\columnwidth]{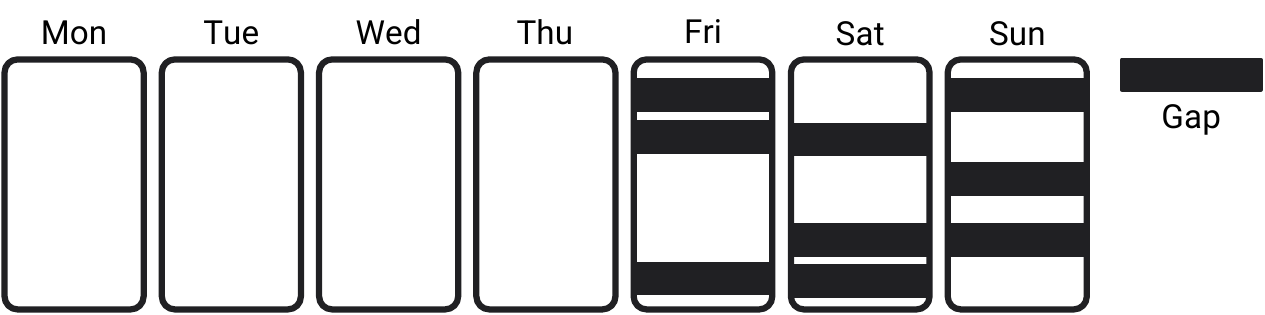}
    \caption{Weekly vehicle schedule illustrating idle gaps (black bars) arising from shipment cancellations. Each column represents one day of the week. These gaps, if left unfilled, directly reduce revenue and fleet utilization. In the present qubit-limited setting of near-term quantum hardware, the weekly problem is made accessible by fixing decisions earlier in the week and exposing variables primarily toward the end; with larger quantum hardware, this variable window could be extended progressively toward the start of the week.}
    \label{fig:gap_schedule_week}
\end{figure}

Maximizing vehicle utilization is a fundamental requirement for operational efficiency in large-scale logistics. For operators of electric freight fleets, maintaining high utilization is complicated by frequent stochastic disruptions, primarily shipment cancellations as well as technical or personnel-related failures. These events create idle ``gaps'' in pre-optimized vehicle schedules that must be remediated to prevent revenue loss and underutilization. \cref{fig:gap_filling_illustration,fig:gap_schedule_week} illustrate this disruption-management setting for electric freight operations.

The class of problems arising from dynamic schedule repair has a long history in operations research. Classical approaches to vehicle routing with time windows use insertion heuristics, branch-and-price, and large neighborhood search as their principal building blocks~\cite{solomon1987algorithms, pillac2013review}. In industrial fleet management, these methods operate in a rolling-horizon loop: when a shipment is canceled, the affected route is repaired by a greedy or local-search procedure that evaluates candidate insertions one vehicle at a time. 
Einride employs a state-of-the-art Electric Vehicle Routing Problem (E-VRP) solver, the Einride Optimization Solver (EOS), in precisely this role, continuously re-optimizing routes under electric vehicle constraints including charging schedules, driver shift limits, and time-window feasibility \cite{einride2024scaling}. While effective at minimizing per-vehicle linear costs, these solvers do not model joint effects between assignments on different vehicles: selecting two sequences on separate vehicle shifts can interact through geographic or temporal overlap in ways that are invisible to per-gap local optimization. The formulation and quantum approach presented in this paper are motivated by this structural gap.

The shipment selection problem (SSP) involves identifying the optimal sequences from a large shipment pool to fill these operational gaps. Each candidate sequence represents a trade-off between economic value and operational risk. While filling a gap with a high-value shipment increases immediate revenue, it may introduce risks such as tight time-window margins or geographic displacement that complicate subsequent assignments. This is particularly critical in electric vehicle (EV) operations, where short-term schedule repairs must remain compatible with long-term constraints, including charging requirements and battery state-of-health.

This study presents a methodology for solving the SSP by balancing these competing objectives within a hybrid quantum-classical workflow. By formulating the task as an optimization problem that accounts for both the marginal benefit of an insertion and its associated risk, we provide a framework for creating robust, high-utilization schedules. The approach ensures that schedule repairs are not only profitable in isolation but also maintain the operational stability necessary for the continuous coordination of electric freight networks.

From a computational perspective, the SSP generalizes NP-hard routing and assignment problems such as the Capacitated Vehicle Routing Problem (CVRP), multi-depot Vehicle Routing Problem (MDVRP), and pickup-and-delivery variants. Recent quantum approaches to related logistics problems have mainly followed two routes. Quantum annealing formulations encode routing and assignment decisions as Quadratic Unconstrained Binary Optimization (QUBO) problems \cite{Neukart2017Traffic, Harikrishnakumar2020MDCVRP}, while hybrid decomposition workflows use classical pre-processing together with quantum subproblem optimization \cite{Stollenwerk2019Portfolio,Feld2019hybrid}. On the gate-model side, QAOA-based approaches have explored tailored encodings and constraint-preserving mixers for routing-type instances \cite{Hadfield2019QAOA}. These studies establish that hybrid quantum-classical methods can be meaningfully applied to logistics optimization, while also highlighting the need for workflows that remain effective under limited circuit depth, approximate simulation, and realistic industrial constraints.

In this work, we develop such a workflow for the shipment selection problem. Specifically, we formulate SSP as a mixed-integer quadratic program and map it to a QUBO and corresponding Ising cost Hamiltonian; we study the resulting optimization problem using noiseless simulations of Iterative-QAOA, a non-variational warm-start extension of QAOA \cite{lopez2025non}; for the scenarios within reach of current devices (20--35 qubits) we corroborate these results with direct execution on IonQ trapped-ion quantum hardware; and we evaluate the resulting assignments within a hybrid EOS warm-start pipeline using operational Key Performance Indicators (KPIs) on real weekly fleet schedules. The rest of the paper is organized as follows. In \cref{sec:problem_formulation} we present the SSP formulation and its Hamiltonian mapping, together with the Iterative-QAOA algorithm. In \cref{sec:methods} we describe the benchmark instances, implementation details, classical baselines, and evaluation protocol. In \cref{sec:results} we report noiseless simulation results, hardware-validation runs, and the hybrid workflow KPIs. Finally, in \cref{sec:future-outlook} we summarize the main findings and discuss future directions. 

\section{Problem Formulation and Quantum Approach}\label{sec:problem_formulation}

\subsection{Mathematical Formulation as a Mixed-Integer Quadratic Program}\label{subsec:miqp}

In this section, we explain how we formulate the SSP for
each gap as a mixed-integer programming model. This framework is commonly used
in optimization and systems research to encode discrete decision-making under
capacity and exclusivity constraints.

Let $G$ denote a finite set of gaps. Each gap $g \in G$ is associated with a
capacity $C_g$. For every gap $g$, let $Q_g$ denote the corresponding set of
feasible sequences. Each sequence $q \in Q_g$ is characterized by a scalar value
$v_{gq}$, a time consumption $t_{gq}$, and a subset of shipments $S_{gq}$ drawn
from a global shipment set $S$.

We introduce binary decision variables
\begin{equation}
x_{gq} =
\begin{cases}
1, & \text{if sequence $q$ is selected for gap $g$}, \\
0, & \text{otherwise},
\end{cases}
\label{eq:decision}
\end{equation}
for all $g \in G$ and $q \in Q_g$.

In shipment selection, quadratic objective terms arise naturally to capture interactions between insertions, such as operational risk, load balancing, or congestion effects; here they encode pairwise schedule compatibility. Accordingly, we maximize
\begin{equation}
    \max \;
    \sum_{g \in G} \sum_{q \in Q_g} v_{gq} x_{gq}
    + \sum_{(q_1,q_2)} w_{q_1 q_2} \, x_{g_1 q_1} x_{g_2 q_2},
    \label{eq:objective}
\end{equation}
where the linear coefficients $v_{gq}$ capture the intrinsic value of individual sequences, and the quadratic term rewards assigning geographically clustered sequences to temporally compatible gaps; pairwise compatibility effects that the per-sequence values alone cannot express.

This structure is closely related to the classical quadratic assignment problem (QAP), in which the cost of assigning facilities to locations is expressed as the product of a flow matrix and a distance matrix. In the present formulation, the coefficients $w_{q_1 q_2}$ play an analogous role, encoding interaction effects induced by assigning sequences $q_1$ and $q_2$ to gaps $g_1$ and $g_2$, respectively.

The optimization is subject to the following constraints:
\textit{Capacity constraints}
enforce feasibility within each gap:
\begin{align}
    \sum_{q \in Q_g} t_{gq} x_{gq} \le C_g,
    \qquad \forall g \in G.
    \label{eq:capacity}
\end{align}
\textit{Shipment exclusivity constraints} ensure that each shipment is used at most once:
\begin{align}
    \sum_{g \in G} \sum_{q \in Q_g : s \in S_{gq}} x_{gq} \le 1,
    \qquad \forall s \in S.
    \label{eq:shipment}
\end{align}
\textit{Gap exclusivity constraints} ensure that at most one sequence is selected for each gap:
\begin{align}
    \sum_{q \in Q_g} x_{gq} \le 1,
    \qquad \forall g \in G.
    \label{eq:gap-exclusivity}
\end{align}
The resulting formulation is a mixed-integer quadratic program (MIQP).

\subsection{From MIQP to Cost Hamiltonian}
\label{subsec:ssc_hamiltonian}

To prepare the SSP for quantum optimization, the constrained MIQP of \cref{subsec:miqp} must be cast as a Hamiltonian energy minimization problem. This is achieved through two successive steps: conversion to a QUBO problem and a standard mapping to an Ising Hamiltonian.

Before encoding, the objective coefficients are jointly normalized. Let
\begin{align}
    V = \sum_{g \in G}\sum_{q\in Q_g} v_{gq},
    \qquad
    W = \sum_{(q_1,q_2)} w_{q_1 q_2},
    \label{eq:coefficients}
\end{align}
and define the normalization factor $\mathcal{N} = (V+W)/100$. All linear and quadratic objective coefficients are rescaled as $v_{gq}\leftarrow v_{gq}/\mathcal{N}$ and $w_{q_1 q_2}\leftarrow w_{q_1 q_2}/\mathcal{N}$, so that the rescaled objective lies in a consistent numerical range across instances of different sizes. The pairwise interaction coefficients $w_{q_1 q_2}$ are generated from a flow matrix $f_{q_1 q_2}$ over sequences and a distance matrix $d_{g_1 g_2}$ over gaps via
\begin{align}
    w_{q_1 q_2} = \lambda_Q\, f_{q_1 q_2}\, d_{g(q_1)\,g(q_2)},
    \label{eq:interaction}
\end{align}
where $g(q)$ denotes the (unique) gap to which sequence $q$ belongs and $\lambda_Q$ is a scaling hyperparameter. To avoid double counting, the implementation includes each unordered pair of sequences at most once; if the input flow matrix is not symmetric, this convention effectively selects one direction.

The shipment exclusivity constraints \cref{eq:shipment} and gap exclusivity constraints \cref{eq:gap-exclusivity} are enforced through quadratic penalties in the QUBO objective using a penalty parameter $\lambda>0$, yielding the unconstrained cost function
\begin{align}
    \nonumber
    C(\mathbf{x}) & = - \left[
        \sum_{g,q} v_{gq} x_{gq}
        + \sum_{(q_1,q_2)}\! w_{q_1 q_2}\, x_{g_1 q_1} x_{g_2 q_2}
    \right]\\
    & + \lambda \sum_{s \in S} P_s(\mathbf{x})
    + \lambda \sum_{g \in G} P_g(\mathbf{x}),
    \label{eq:qubo}
\end{align}
where $P_s(\mathbf{x})$ and $P_g(\mathbf{x})$ are quadratic penalty terms that vanish if and only if the respective constraint is satisfied. The sign convention follows the QUBO minimization framework: the negated and normalized objective is minimized, so that lower-energy solutions correspond to higher-utility assignments. The capacity constraints \cref{eq:capacity} are not penalized in the QUBO; instead, violations are detected from measurement outcomes and handled as a soft penalty during the post-processing stage of energy evaluation (see \cref{sec:application}), avoiding the slack variables (qubits) that a direct QUBO encoding of \cref{eq:capacity} would require.

Given $C(\mathbf{x})$, the cost Hamiltonian $H_C$ acting on $n = \sum_{g \in G}|Q_g|$ qubits is constructed by promoting each binary variable to a quantum operator via the standard substitution $x_{gq} \to (I - Z_{gq})/2$, where $Z_{gq}$ is the Pauli-$Z$ operator on the qubit associated with $x_{gq}$. Applying this substitution to \cref{eq:qubo} yields a diagonal Hamiltonian of the Ising form
\begin{align}
    \nonumber
    H_C &= E_0\,I + \sum_{(g,q)} h_{gq}\, Z_{gq}\\
          &+ \sum_{(g_1,q_1) < (g_2,q_2)}\!\! J_{(g_1 q_1)(g_2 q_2)}\, Z_{g_1 q_1} Z_{g_2 q_2},
    \label{eq:hamiltonian}
\end{align}
where $E_0$ is a constant energy offset, $h_{gq}$ are local fields arising from the linear terms and the diagonal penalty contributions, and $J_{(g_1 q_1)(g_2 q_2)}$ are two-body couplings arising from both the quadratic objective and the off-diagonal penalty terms.

\subsection{Iterative-QAOA}
\label{sec:iterqaoa}

With the cost Hamiltonian $H_C$ established, the optimization problem reduces to finding low-energy bit strings of $H_C$, whose ground state encodes an optimal (or near-optimal) SSP assignment. We address this using the Iterative-QAOA algorithm \cite{lopez2025non}, a non-variational extension of QAOA that replaces classical parameter optimization with a fixed parameter schedule and an iterative, measurement-driven warm start.

Iterative-QAOA uses the standard depth-$p$ QAOA ansatz
\begin{align}
    \ket{\psi_p(\boldsymbol{\gamma},\boldsymbol{\beta})}
    =
    \prod_{k=1}^{p} e^{-i\beta_k H_M} e^{-i\gamma_k H_C}\,
    \ket{\psi_{\mathrm{init}}},
\end{align}
where $H_M$ is a mixer Hamiltonian and $\ket{\psi_{\mathrm{init}}}$ is a product initial state that is updated across iterations. Rather than optimizing the parameters $(\boldsymbol{\gamma},\boldsymbol{\beta})$ variationally, we employ a linear-ramp (LR-QAOA) schedule \cite{Kremenetski2023-vy, Montanez-Barrera2025-im, Dehn2025-jm} parameterized by a single slope $\Delta$ (the SSP-specific choice of $\Delta$ is discussed in \cref{sec:application}).

After executing the QAOA circuit, measurement outcomes $\{\mathbf{x}_j\}$ are assigned energies $E(\mathbf{x}_j)\equiv \mel{\mathbf{x}_j}{H_C}{\mathbf{x}_j}$ and converted into a Boltzmann distribution
\begin{align}
    P(\mathbf{x}_i) = \frac{e^{-\beta_T E(\mathbf{x}_i)}}{\sum_j e^{-\beta_T E(\mathbf{x}_j)}},
\end{align}
where $\beta_T$ is a tunable hyperparameter. Using this distribution, Iterative-QAOA computes a qubit-wise bias $m_q = \sum_{j} P(\mathbf{x}_j)(-1)^{z_q}$ and updates the next product-state initialization via
\begin{align}
    \rho_q = \frac{1}{2}\left(1 - \eta m_q\right),
    \qquad \eta\in\{-1,+1\}.
\end{align}
The updated initial state for iteration $j+1$ is then prepared as
\begin{align}
    |\psi_{\mathrm{init}}^{(j+1)}\rangle = \bigotimes_{q=1}^{n}\left(\sqrt{1 - \rho_q}\ket{0}_q+\sqrt{\rho_q}\ket{1}_q\right),
    \label{eq:init_state}
\end{align}
using single-qubit $R_y$ rotations. Following the warm-start QAOA construction, the mixer is modified so that $|\psi_{\mathrm{init}}^{(j+1)}\rangle$ is an eigenstate of the mixer Hamiltonian \cite{Egger2021-kw}.

By iterating this procedure the algorithm progressively biases the search toward regions of the Hilbert space associated with low-energy solutions.

\section{Methods}\label{sec:methods}

\subsection{Problem Instances}\label{sec:instances}

The benchmark set comprises instances constructed from anonymized customer data obtained from active electric freight operations managed through Einride's Saga AI platform \cite{einride2026saga}. All shipment identifiers, route details, and facility locations were anonymized before use.

The main benchmark cohort consists of eight real-world weekly schedules drawn from operations spanning more than one year and denoted by the corresponding month labels \textit{Feb, Mar, May, Jun, Jul, Oct, Nov,} and \textit{Dec}. Each weekly schedule covers a one-week planning horizon extracted from continuous fleet operations and serves as the basis for end-to-end validation of the hybrid workflow.

Instance generation follows a decomposition procedure that mimics operational disruptions. Starting from a baseline schedule produced by Einride's E-VRP solver \cite{einride2024scaling}, a stochastic cancellation \textit{scenario} removes a subset of committed shipments at a simulated ``current time'' timestamp. In this study, all weekly schedules are evaluated under the \emph{c11} cancellation level, corresponding to 11 randomly removed shipments, which represents an operationally realistic disruption volume. For each weekly schedule, $N$ independent cancellation scenarios are sampled to capture variability in the disrupted shipment set; the statistics reported in \cref{tab:instances} are aggregated over those $N$ draws.

For every resulting idle gap with a minimum duration of 2 hours, a heuristic enumerates all feasible candidate shipment sequences, filtering against vehicle charging constraints, driving-time regulations, time windows, and driver shift limits. Each candidate sequence is then evaluated by Einride's VRP solver to compute the marginal cost saving $v_{gq}$, defined as the difference in total plan cost between the baseline plan and the plan obtained by inserting sequence $q$ into gap $g$ (see \cref{subsec:miqp}). Because all vehicles are already operational before any insertion, fixed costs cancel identically, and the resulting difference captures variable cost savings in distance, energy, and driver time directly. The resulting scenario instance is defined by the identified gaps and their feasible sequences and is then encoded as the MIQP of \cref{subsec:miqp}. While the model supports multi-sequence extensions, the experiments in this work focus on single-shipment insertions per gap to reflect current industrial requirements.

\begin{table}[t]
    \caption{Weekly benchmark schedules evaluated under the \emph{c11} cancellation level (11 randomly removed shipments). $|G|$ denotes the average number of idle gaps across the $N$ sampled scenarios for a given weekly schedule, $n_\text{max}$ is the maximum number of binary variables (qubits) observed across those scenarios, $|\overline Q_g|$ is the mean number of candidate sequences per gap, and $N$ is the number of independently sampled cancellation scenarios.}
    \label{tab:instances}
    \centering
    \small
    \begin{tabular*}{\columnwidth}{@{\extracolsep{\fill}}lrrrr}
    \toprule
     Instance & $|G|$ & $n_\text{max}$ & $|\overline Q_g|$ & $N$ \\
    \midrule
    Feb & 6.7 & 42 & 3.2 & 3 \\
    Mar & 8.6 & 55 & 4.5 & 6 \\
    May & 13.7 & 90 & 5.0 & 4 \\
    Jun & 10.3 & 119 & 8.9 & 4 \\
    Jul & 8.7 & 130 & 15.0 & 6 \\
    Oct & 10.0 & 94 & 7.3 & 3 \\
    Nov & 9.0 & 74 & 6.9 & 3 \\
    Dec & 8.7 & 72 & 7.2 & 3 \\
    \bottomrule
    \end{tabular*}
\end{table}

\subsection{Application of Iterative-QAOA to the Shipment Selection Problem}
\label{sec:application}

Each binary decision variable $x_{gq}$ is encoded into one qubit, yielding a QAOA circuit of width $n = \sum_{g \in G} |Q_g|$. We incorporate the shipment exclusivity constraint \cref{eq:shipment} and the gap exclusivity constraint \cref{eq:gap-exclusivity} directly in the QUBO via quadratic penalties, so no slack qubits are required. Unless stated otherwise, we set the penalty weight to $\lambda=10$.

The capacity constraints \cref{eq:capacity} are not embedded in the QUBO. Instead, after each measurement, every bitstring is decoded into an assignment $\mathbf{x}$ and checked for capacity feasibility per gap. Defining the overload $\Delta_g(\mathbf{x}) = \sum_{q \in Q_g} t_{gq}\,x_{gq} - C_g$, a smooth saturating penalty
\begin{align}
    E_{\mathrm{cap}}(\mathbf{x})
    = \lambda
      \sum_{g \in G}
      \mathrm{erf}\!\left(\max\!\left(0,\,\Delta_g(\mathbf{x})\right)\right),
    \label{eq:capacity_penalty}
\end{align}
with $\lambda = 10$, is added to the Hamiltonian energy before ranking samples and updating the warm-start bias. The $\mathrm{erf}$ form grows smoothly with the violation magnitude and saturates at large overloads, preventing infeasible solutions from dominating the energy scale.

Each QAOA circuit uses depth $p=5$ for instances with $n<50$ qubits and $p=6$ otherwise. The cost layer implements the unitary $\exp(-i\gamma_k H_C)$ via single-qubit $R_z$ rotations for one-body $Z$ terms and entangling $R_{zz}$ rotations for two-body $ZZ$ couplings. For realistic instances, the number of $ZZ$ terms can be large. To control the two-qubit gate count, we apply a heuristic pruning scheme: couplings are sorted by decreasing magnitude and only the top $\lfloor 30\,n/p \rfloor$ per layer are retained; any coupling falling below $1\%$ of the maximum is discarded. All local $Z$ terms are kept in full. The QAOA ansatz is built from this pruned Hamiltonian operator, while energies and constraint violations (including $E_{\mathrm{cap}}$) are evaluated with the full Hamiltonian $H_C$.

Because pruning caps the retained couplings at $\lfloor 30n/p \rfloor$ per layer, the two-qubit gate count is $O(n)$: the largest simulated circuit ($n=130$, $p=6$) is composed of $7{,}800$ two-qubit and $7{,}150$ single-qubit gates, and a hardware-scale instance ($n=35$, $p=5$) of $2{,}100$ two-qubit and $1{,}785$ single-qubit gates, compiling, after circuit optimization, to $819$ $R_{zz}$ and $412$ $\mathrm{GPI2}$ native gates. Since this count is effectively independent of $p$, the deeper $p=6$ ansatz adopted for the larger instances, which we heuristically expect to benefit from a more expressive circuit, incurs little additional two-qubit cost.

\begin{figure}[t!]
    \centering
    \includegraphics[width=\columnwidth]{
    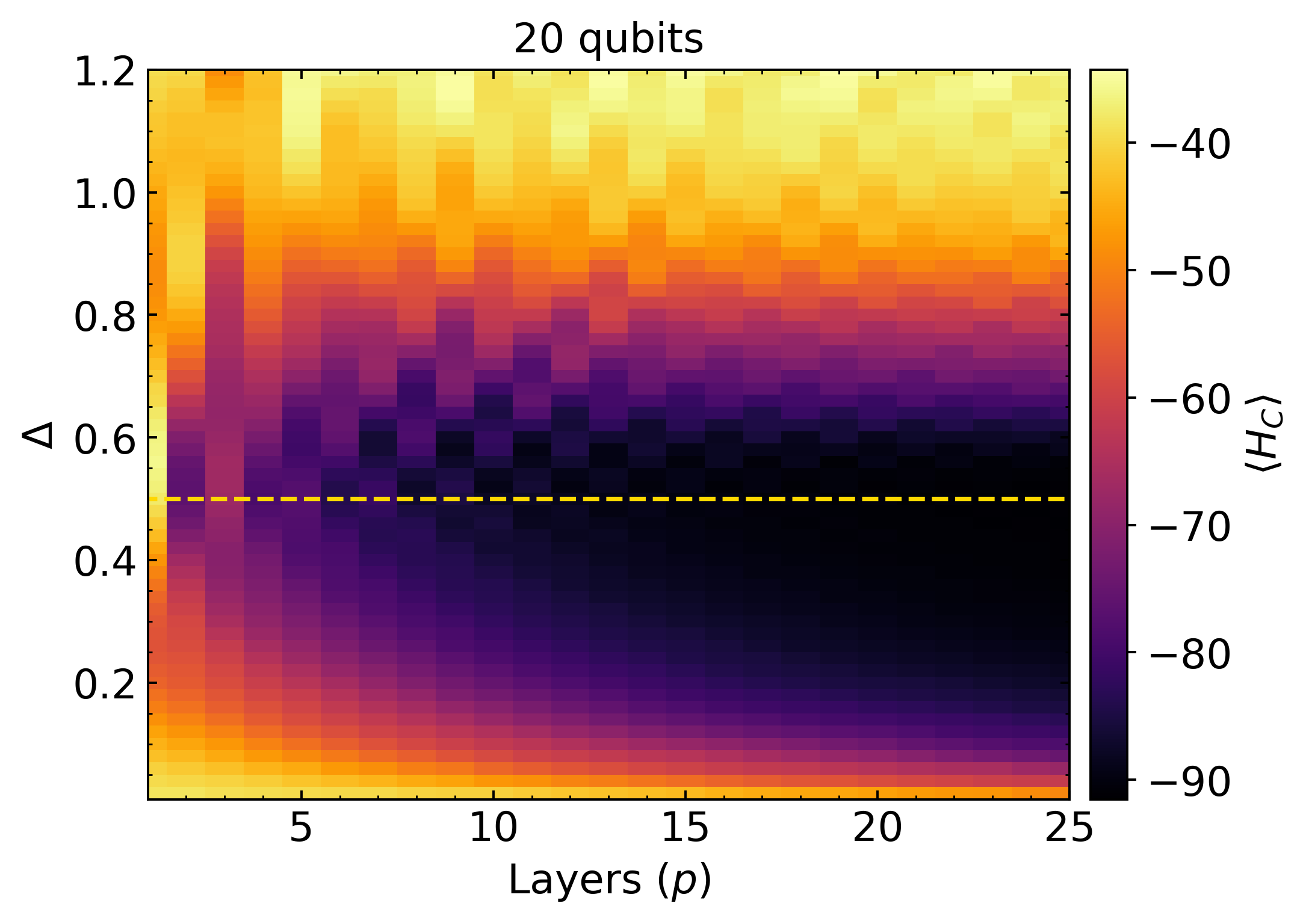}
    \caption{LR-QAOA performance landscape for a representative 20-qubit SSP instance. The heatmap shows the expectation value of the cost Hamiltonian $\ev{H_C}$ as a function of the slope parameter $\Delta$ and circuit depth $p$. The dashed horizontal line indicates the instance-optimal $\Delta$ that minimizes $\ev{H_C}$.}
    \label{fig:lrqaoa_param_exp}
\end{figure}

\begin{figure}[t!]
    \centering
    \includegraphics[width=\columnwidth]{
    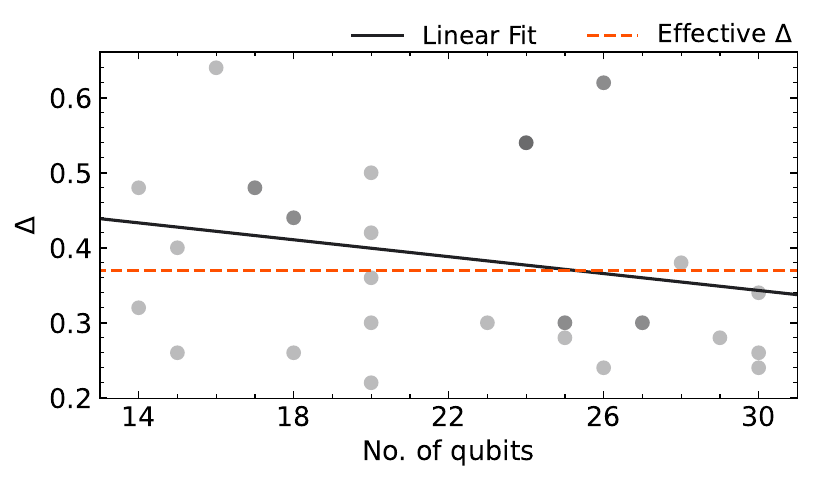}
    \caption{Optimal LR-QAOA parameter $\Delta$ across 32 SSP instances of varying sizes. Each point marks the value of $\Delta$ that minimizes $\ev{H_C}$ for a single instance; point opacity is proportional to the number of instances sharing that value, so darker points indicate a higher incidence. The solid line is a linear fit to the data. The dashed line indicates $\Delta = 0.37$, used for instances with $n<30$ qubits. Due to the observed decreasing trend with instance size, we use a smaller heuristic value $\Delta = 0.30$ for $n\ge 30$ qubits (not shown).}
    \label{fig:lrqaoa_deltas}
\end{figure}

Angles are assigned using the LR-QAOA schedule
\begin{align}
    \gamma_k = \frac{k}{p}\Delta, \qquad
    \beta_k  = \frac{p-k+1}{p}\Delta, \qquad k=1,\dots,p,
    \label{eq:lr_schedule}
\end{align}
so that a single slope parameter $\Delta$ defines the full parameter set. Based on noiseless simulation sweeps (see \cref{fig:lrqaoa_param_exp,fig:lrqaoa_deltas}) we observe that the typical optimal $\Delta$ decreases mildly with system size. We therefore use the heuristic choice $\Delta=0.37$ for $n<30$ and $\Delta=0.30$ for $n\ge 30$.

The warm-starting feedback loop runs for $N_\text{iter}=20$ iterations, where each circuit is executed with $4{,}000$ shots. For each measured bitstring, the full energy $E(\mathbf{x})=\langle \mathbf{x}|H_C|\mathbf{x}\rangle$ is computed and, if capacity constraints are violated, the soft penalty \cref{eq:capacity_penalty} is added. In the subsequent warm-start update we use the 100 lowest-energy samples.
The inverse temperature hyperparameter $\beta_T$ is chosen as follows. Let $E_1\le E_2\le\dots$ denote the selected energies, shifted so that $E_1=0$. We compute the adjacent gaps $\Delta E_m = E_{m+1}-E_m$ and determine an iteration-dependent $\beta_T$ by rescaling a base quadratic schedule,
\begin{align}
    \beta_T(j) = 0.1 + 0.9\left(\frac{j}{N_\text{iter}}\right)^2 \qquad j = 0, 1, \dots, N_\text{iter}-1,
\end{align}
using the smallest energy gap that exceeds a tolerance threshold $\tau$:
\begin{align}
    \beta_T(j) \leftarrow \frac{\beta_T(j)}{\min\{\Delta E_m:\Delta E_m>\tau\}}.
    \label{eq:beta_rescale}
\end{align}
If no $\Delta E_m$ exceeds $\tau$, we re-use the previous $\beta_T$. In practice, we take $\tau$ to decay with iteration, $\tau(j) = 0.1/y(j)^2$ where $y(j)$ linearly interpolates from $1$ to $\sqrt{10}$ across $N_\text{iter}+1$ points, so the tolerance runs from $\tau(0) = 0.1$ down to $\approx 0.01$ at the final iteration. This rescaling is an implementation heuristic that adapts the sharpness of the Boltzmann weights to the observed energy spacing in the low-energy tail; the specific constants are not finely tuned, and the method does not depend on their precise values. A simpler fixed $\beta_T$ would still work but risks weighting samples too uniformly or collapsing onto a single one too early, so the adaptive rule mainly speeds early convergence rather than changing final quality.
The initial state  $|\psi_{\mathrm{init}}^{(j)}\rangle$ (\cref{eq:init_state}) at iteration $j$ is prepared by single-qubit $R_y(\theta_q)$ rotations with angles $\theta_q = 2\arcsin\sqrt{\rho_q}$, and $\rho_q$ is clipped to $[10^{-3},1-10^{-3}]$. To ensure that the warm-start state is compatible with the mixing Hamiltonian, each mixer unitary is transformed as a $R_z$ rotation \cite{Egger2021-kw}:
\begin{align}
    R_y(-\theta_q)\,R_z(-2\beta_k)\,R_y(\theta_q).
\end{align}

After the Iterative-QAOA run, the resulting candidate quantum solutions are post-processed with the in-house refinement heuristic described in \cref{sec:refinement}. We then retain the ten lowest-energy refined solutions for each instance and use this batch as input to the reconstruction and warm-start evaluation pipeline described in \cref{sec:merit-factor-eval}.

While the principal results are obtained from noiseless simulation, we additionally execute Iterative-QAOA on IonQ Forte-generation trapped-ion quantum hardware for the subset of scenarios within current device reach, spanning 20--35 qubits. These hardware runs use the identical algorithm configuration as the corresponding simulations and use only IonQ's default debiasing setting, with no further error mitigation, providing a direct test of whether the simulated workflow transfers to hardware. Device characteristics are summarized in \cref{sec:SI_hardware_details}, and an instance-level hardware-versus-simulation comparison is reported in \cref{sec:SI_hardware}.

\subsection{Classical Solvers}

We use two classical procedures for distinct purposes. SCIP provides the classical reference used to benchmark Iterative-QAOA on the MIQP defined in \cref{sec:problem_formulation}. Separately, an in-house refinement heuristic is applied as a lightweight post-processing step to the quantum solutions before the final candidate batch is selected for downstream reconstruction and KPI evaluation.

\subsubsection{SCIP for baseline investigation}

To establish a classical baseline for comparison with Iterative-QAOA, we solve the MIQP of \cref{sec:problem_formulation} using the SCIP optimization suite \cite{bolusani2024scip}. SCIP is a state-of-the-art mixed-integer solver with native support for quadratic objectives and is therefore well suited to the QAP-like structure of our formulation. In this work, SCIP is used only as an objective-space baseline for the quantum solver; the operational workflow baseline used for KPI evaluation is EOS, as described in \cref{sec:merit-factor-eval}.

We access SCIP through PySCIPOpt and represent the quadratic interaction term explicitly through the compatibility matrix defined by the product of the distance matrix over gaps and the flow matrix over sequences. This preserves the assignment-dependent structure of the SSP and enables direct comparison with the cost-Hamiltonian values optimized by Iterative-QAOA. Throughout the solution process, we record the best incumbent objective value together with standard solver diagnostics such as primal and dual bounds, mixed-integer optimality gap, and processed node counts.

\subsubsection{In-House refinement heuristic}\label{sec:refinement}

In addition to the SCIP baseline, we use an in-house refinement heuristic as a lightweight post-processing step for the candidate solutions returned by Iterative-QAOA. Starting from a feasible assignment, the heuristic performs a small number of feasibility-preserving local moves that add or remove a single sequence and keeps the best improved solution encountered. Its purpose is to make inexpensive local corrections to raw quantum outputs before the final refined candidate batch is selected for downstream reconstruction and KPI evaluation. Full implementation details and pseudocode are deferred to the Supplementary Information (\cref{sec:SI_in_house_refinement}, \cref{alg:inhouse_refinement}).

\subsection{Evaluation Metrics and Hybrid Workflow}\label{sec:merit-factor-eval}

\begin{figure}[t!]
    \centering
    \includegraphics[width=\columnwidth]{
    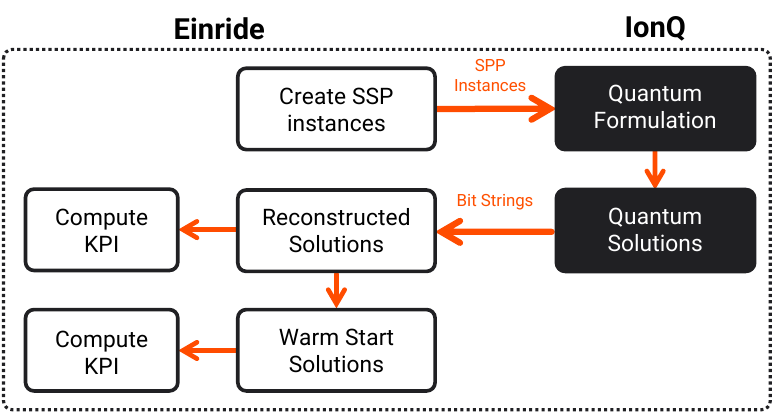}
    \caption{Overview of the hybrid 3-step evaluation process between Einride and IonQ.}
    \label{fig:workflow}
\end{figure}

We evaluate the hybrid workflow on a scenario-matched basis using the three-step process shown in \cref{fig:workflow}. For each weekly schedule and each cancellation scenario, we compare three operational solutions produced from the same disrupted instance: (i)~the cold-start Einride Optimization Solver (EOS) baseline, denoted \emph{BL}; (ii)~the \emph{reconstructed quantum} solution, denoted \emph{Q}, obtained by inserting the selected quantum assignment directly into the vehicle plans without further EOS re-optimization; and (iii)~the \emph{quantum warm-start} solution, denoted \emph{QWS}, obtained by passing the reconstructed plan to EOS as an initial solution. Because different cancellation scenarios correspond to different SSP instances, all KPI comparisons are performed within the same scenario and aggregated only afterward.

\Cref{fig:workflow} also clarifies the role of each stage in the hybrid pipeline. The comparison between BL and Q isolates the direct effect of the quantum assignment, since Q is obtained by reconstructing the Iterative-QAOA selection without any additional classical re-optimization. The comparison between Q and QWS then isolates the contribution of the warm-start step, in which EOS receives the reconstructed quantum plan as an initial solution and performs a bounded refinement pass. In this way, the three-way comparison distinguishes what is gained from the quantum selection itself from what is gained by embedding that selection into the downstream classical optimization workflow.

The primary KPI used to measure delivered service is \emph{Shipments Delivered} (SD), defined as the number of shipments included in feasible vehicle routes after reconstruction or warm-start refinement. We also track \emph{Total Operational Cost} (TC),
\begin{equation}
  C_\text{ops} :=  \sum_{s} c^\text{fleet}_s + \sum_{r} c^\text{reject}_r ,
\end{equation}
which aggregates fleet-execution cost terms $c^\text{fleet}_s$ for delivered shipments $s$ and rejection penalties $c^\text{reject}_r$ for undelivered shipments $r$. This quantity is aligned with the linear contribution to the MIQP objective: the coefficient $v_{gq}$ represents the cost saving $\Delta C_{\text{ops},qg}$ associated with inserting sequence $q$ into gap $g$ relative to the baseline plan. In the main results, however, we focus on SD together with routing-efficiency and compatibility metrics, since these most directly expose the effect of the quantum assignment and the subsequent warm-start.

Routing efficiency is quantified through \emph{Total Drive Distance} (TDD) and \emph{Drive Distance per Shipment} (TDD/SD). The pairwise quality of the selected assignment is quantified by the \emph{Schedule Compatibility Score} (SCS), defined for $k = \sum_{g,q} x_{gq}$ selected sequences as
\begin{align}
    \mathrm{SCS}(\mathbf{x})
    \;=\; \frac{1}{k^2} \sum_{(q_1,q_2)} w_{q_1 q_2}\, x_{g_1 q_1}\, x_{g_2 q_2},
    \label{eq:scs}
\end{align}
where $w_{q_1 q_2}$ is given in \cref{eq:interaction}. Higher SCS indicates that the selected insertions are more geographically co-located and temporally compatible, which may have longer-term operational benefits through battery charging, maintenance, and vehicle load balancing.

\section{Results}\label{sec:results}

\subsection{Results of noiseless Iterative-QAOA simulations}\label{sec:noiseless_sims}

\begin{figure*}[t!]
    \centering
    \includegraphics[width=\textwidth]{
    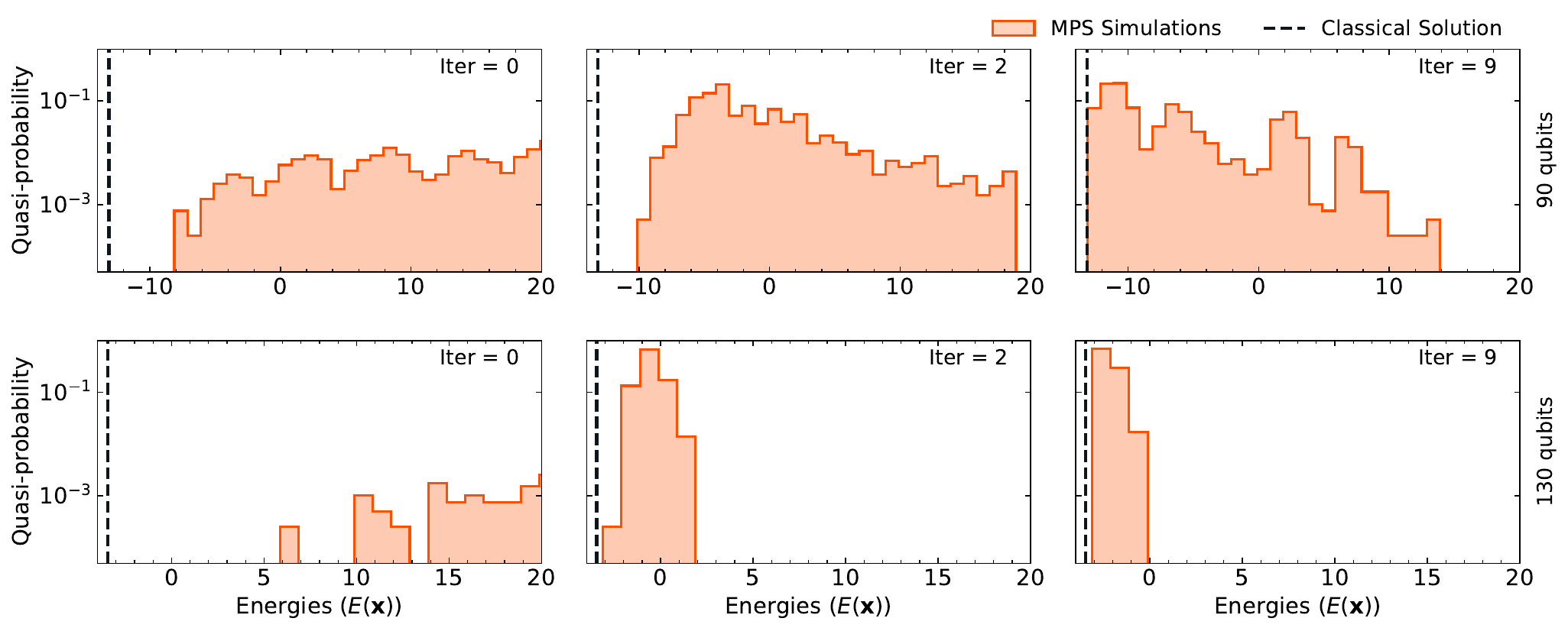}
    \caption{Iterative-QAOA executed on two large SSP instances using an MPS simulator with bond dimension $\chi = 256$. The top row corresponds to a May 90-qubit scenario and the bottom row to a July 130-qubit scenario. For both instances, $\Delta = 0.30$ with QAOA circuits of depth $p = 6$. Each column reports the sampled quasi-probability distribution over cost Hamiltonian energies at iterations $\mathrm{Iter}=0,2,9$. The vertical dashed line indicates the baseline solution obtained by the SCIP classical solver.}
    \label{fig:iterqaoa_mps-ideal}
\end{figure*}

We carried out noiseless simulations of Iterative-QAOA across the full benchmark set, using a statevector simulator for $n \le 32$ qubits and a Matrix Product State (MPS) simulator with bond dimension $\chi=256$ for larger instances. \cref{fig:iterqaoa_mps-ideal} shows representative large-instance results for May (90 qubits) and July (130 qubits); since the optimization is a ground-state search, the horizontal axis reports sampled cost-Hamiltonian energies rather than operational objective values. The figure therefore gauges Iterative-QAOA convergence directly. At initialization the sampled distribution is broad, as the fixed linear-ramp schedule explores a wide portion of the feasible search space. After only a few warm-start updates it shifts toward smaller energies with the high-energy tail strongly suppressed, and by $\mathrm{Iter}=9$ the weight is concentrated in the low-energy sector, with further iterations no longer changing it meaningfully.

The sharpness of this contraction is instance dependent. For May, the final distribution remains spread across several nearby low-energy bins, indicating multiple competitive assignments, whereas July contracts markedly, with most probability mass near the lowest energies by $\mathrm{Iter}=9$.

At these instance sizes, the MPS approximation is expected to remain accurate given the circuits are relatively shallow $(p\leq6)$, the pruned Hamiltonian is sparse, and the warm start biases the state toward product-like configurations, keeping entanglement modest. Moreover, as simulation cost grows rapidly with bond dimension, $\chi=256$ is a trade-off between accuracy and computational time, with truncation error remaining a factor for the largest instances.

A broader summary of solution quality across the full benchmark set, including the effect of the classical refinement procedure of \cref{sec:refinement}, is given in the Supplementary Information (\cref{sec:SI_iterqaoa_performance}).

For each instance, the candidate solutions generated by Iterative-QAOA are first refined using the in-house heuristic of \cref{sec:refinement}. We then retain the ten lowest-energy refined solutions, producing a batch of quantum solutions $\{\mathbf{x}^{*}\}$ that is subsequently reconstructed and evaluated as described in \cref{sec:merit-factor-eval}.

\begin{figure*}[t!]
    \centering
    \includegraphics[width=\textwidth]{
    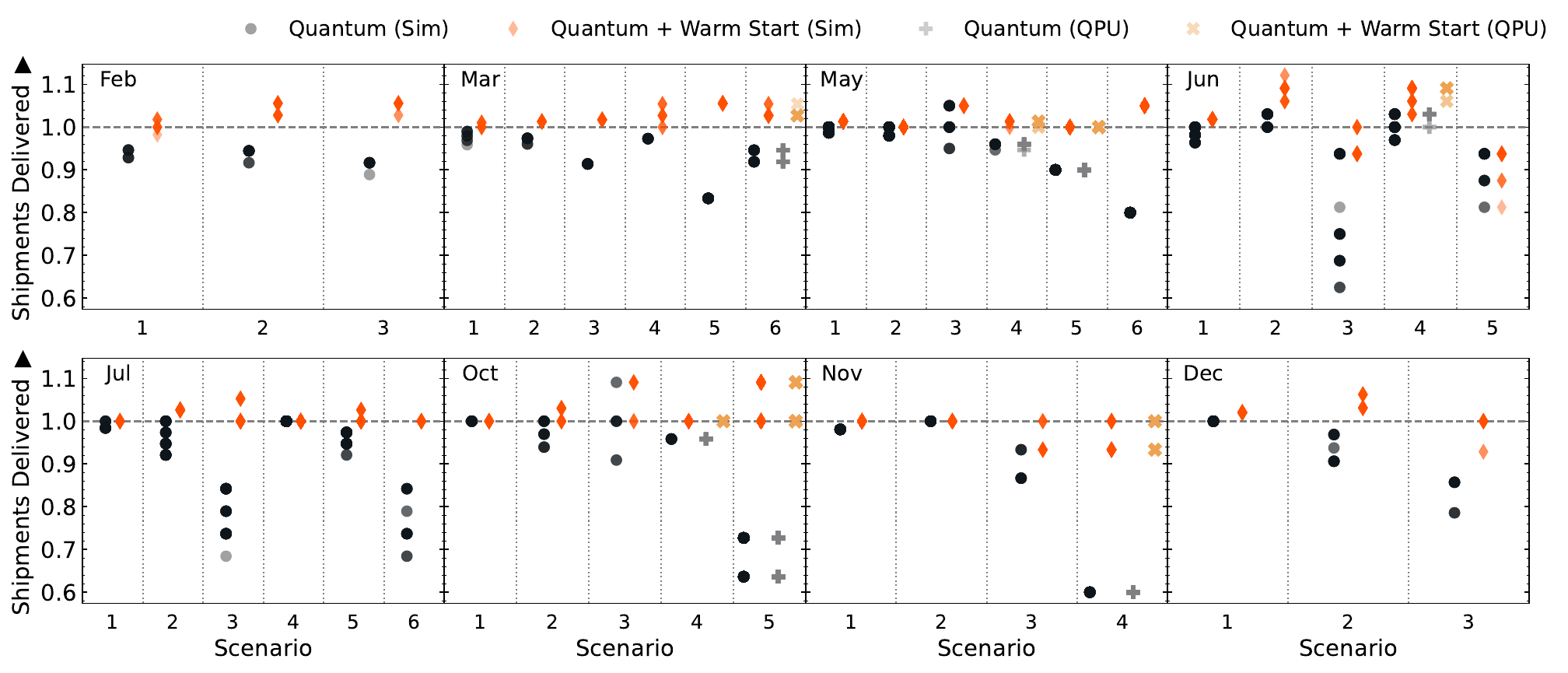}
    \caption{Scenario-level comparison of shipments delivered (SD) relative to the baseline EOS solution. Each point corresponds to one cancellation scenario from one of the eight weekly schedules. The plotted quantity is the ratio $\mathrm{SD}/\mathrm{SD}_{\mathrm{BL}}$ for the reconstructed quantum solution and the quantum warm-start solution, where $\mathrm{SD}_{\mathrm{BL}}$ is the shipments-delivered value of the corresponding baseline (BL) solution for the same scenario. Values above 1 indicate an improvement over the baseline. Markers labeled ``(QPU)'' indicate the scenarios additionally executed on IonQ trapped-ion quantum hardware.}
    \label{fig:kpi_sd}
\end{figure*}

\begin{figure*}[t!]
    \centering
    \includegraphics[width=\textwidth]{
    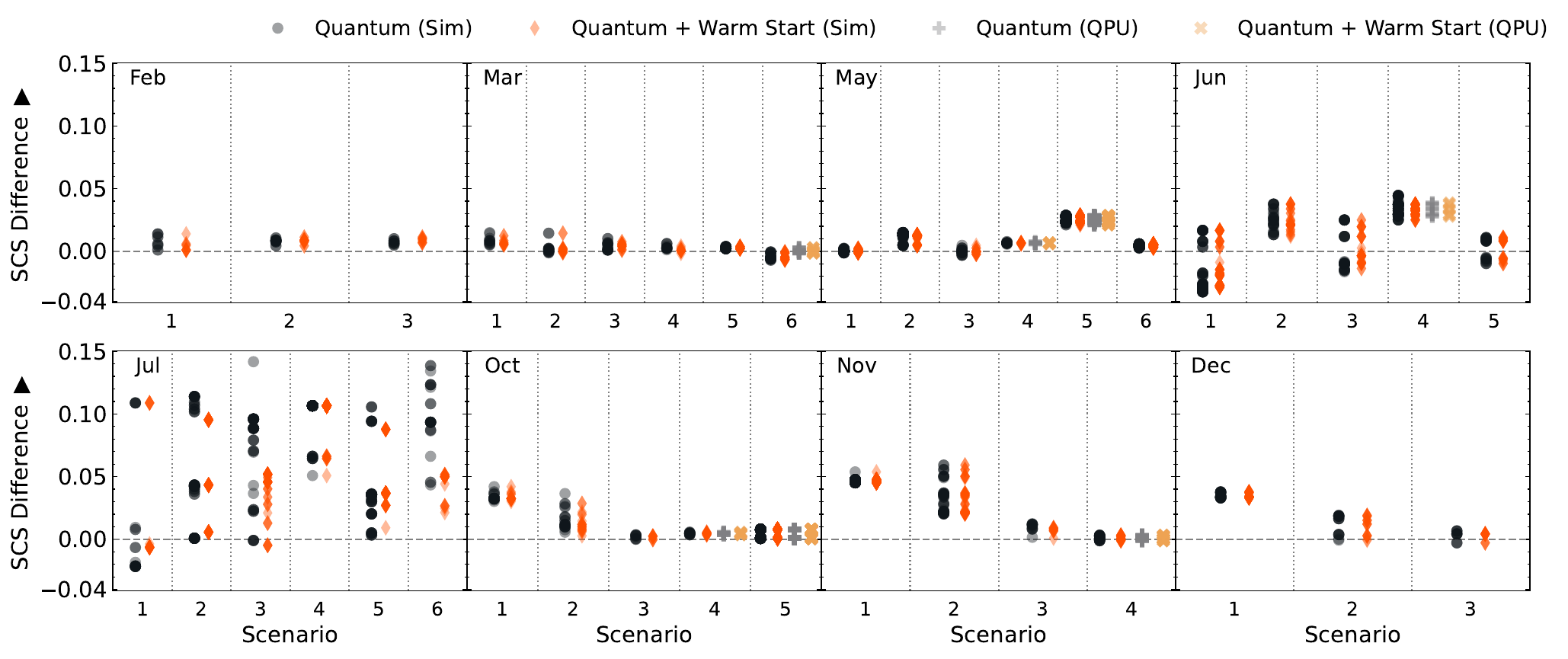}
    \caption{Scenario-level comparison of schedule compatibility score (SCS) relative to the baseline EOS solution. Each point corresponds to one cancellation scenario from one of the eight weekly schedules. The plotted quantity is the difference $\mathrm{SCS}-\mathrm{SCS}_{\mathrm{BL}}$ for the reconstructed quantum solution and the quantum warm-start solution, where $\mathrm{SCS}_{\mathrm{BL}}$ is the baseline value for the same scenario. Positive values indicate improved pairwise schedule compatibility relative to the baseline. Markers labeled ``(QPU)'' indicate the scenarios additionally executed on IonQ trapped-ion quantum hardware.}
    \label{fig:kpi_scs}
\end{figure*}

We present application-level performance for the hybrid workflow. The comparison is scenario matched: for each disrupted weekly-schedule scenario we compare the cold-start EOS baseline (BL), the reconstructed quantum solution (Q), and the quantum warm-start solution (QWS) produced from the same underlying instance, aggregating only afterward. \cref{tab:kpi_avg_improvements} reports these results per month in \emph{average (best)} form (see caption for notation), and the scatter plots in \cref{fig:kpi_sd,fig:kpi_scs} show the underlying scenario-level SD ratio and $\Delta$SCS values directly. The averages summarize the typical behavior of the reconstruction and warm-start pipeline, while the best values highlight the strongest benefit the workflow can deliver on each KPI; together they distinguish the direct contribution of the quantum assignment from the additional gain obtained once EOS refines it as a warm start.

\begin{table*}[t]
    \caption{Scenario-level KPI improvements relative to the baseline EOS solution for the eight weekly schedules. For each cancellation scenario, the reconstructed quantum solution (Q) and the quantum warm-start solution (QWS) are compared against the corresponding cold-start baseline (BL), and the reported values are aggregated over the $N$ scenarios associated with each month. Each entry is written as \emph{average (best)}, where the value outside parentheses is the average improvement over the scenarios for that month and the value in parentheses is the strongest scenario-level improvement obtained for the same method and KPI. The entries for SD, TDD, and TDD/SD are percentage improvements with respect to BL, whereas $\Delta$SCS reports the absolute change in Schedule Compatibility Score. Positive values in SD and $\Delta$SCS indicate improvement over the baseline, while negative values in TDD and TDD/SD indicate more efficient routing. The final row gives the mean over all 32 scenarios, with the corresponding mean best value in parentheses.}
    \label{tab:kpi_avg_improvements}
    \centering
    \small
    \begin{tabular*}{\textwidth}{@{\extracolsep{\fill}}l r rr rr rr rr}
        \toprule
        & & \multicolumn{2}{c}{SD (\%)} & \multicolumn{2}{c}{$\Delta$SCS} & \multicolumn{2}{c}{TDD (\%)} & \multicolumn{2}{c}{TDD/SD (\%)} \\
        \cmidrule(lr){3-4}\cmidrule(lr){5-6}\cmidrule(lr){7-8}\cmidrule(lr){9-10}
        Month & $N$ & Q (Best) & QWS (Best) & Q (Best) & QWS (Best) & Q (Best) & QWS (Best) & Q (Best) & QWS (Best) \\
        \midrule
        Feb & 3 & -6.8 (-5.4) & +3.6 (+5.6) & +0.01 (+0.01) & +0.01 (+0.01) & -5.5 (-8.7) & +0.5 (-1.6) & +1.3 (-1.3) & -2.9 (-5.9) \\
        Mar & 6 & -8.5 (-1.0) & +3.1 (+5.6) & +0.00 (+0.01) & +0.00 (+0.01) & -0.7 (-6.5) & +5.8 (-1.1) & +9.0 (-1.0) & +2.6 (-2.4) \\
        May & 4 & -8.8 (+5.0) & +2.8 (+5.0) & +0.00 (+0.02) & +0.00 (+0.01) & -11.2 (-25.4) & +0.5 (-0.5) & -2.9 (-6.8) & -2.3 (-4.7) \\
        Jun & 4 & -5.5 (+3.0) & +0.3 (+12.1) & -0.00 (+0.04) & -0.00 (+0.04) & -1.9 (-19.5) & +1.1 (-9.4) & +4.4 (-1.8) & +1.0 (-4.0) \\
        Jul & 6 & -3.6 (+0.0) & +0.4 (+5.3) & +0.08 (+0.14) & +0.08 (+0.11) & -3.0 (-32.5) & +1.3 (-2.3) & +0.7 (-3.8) & +0.9 (-4.0) \\
        Oct & 3 & -1.1 (+9.1) & +1.7 (+9.1) & +0.02 (+0.04) & +0.02 (+0.04) & -2.1 (-6.3) & -0.0 (-2.8) & -1.0 (-4.4) & -1.7 (-5.5) \\
        Nov & 3 & -2.1 (+0.0) & -0.3 (+0.0) & +0.04 (+0.06) & +0.04 (+0.06) & -1.2 (-8.5) & +0.6 (-1.3) & +1.0 (-1.3) & +1.0 (-1.3) \\
        Dec & 3 & -4.6 (+0.0) & +2.1 (+6.2) & +0.02 (+0.04) & +0.02 (+0.04) & -2.2 (-13.9) & +1.8 (-3.3) & +2.7 (+0.5) & -0.3 (-3.7) \\
        \midrule
        Mean & - & -5.1 (+1.3) & +1.7 (+6.1) & +0.02 (+0.05) & +0.02 (+0.04) & -3.5 (-15.2) & +1.5 (-2.8) & +1.9 (-2.5) & -0.2 (-3.9) \\
        \bottomrule
    \end{tabular*}%
\end{table*}

\begin{figure}[t!]
    \centering
    \includegraphics[width=\columnwidth]{
    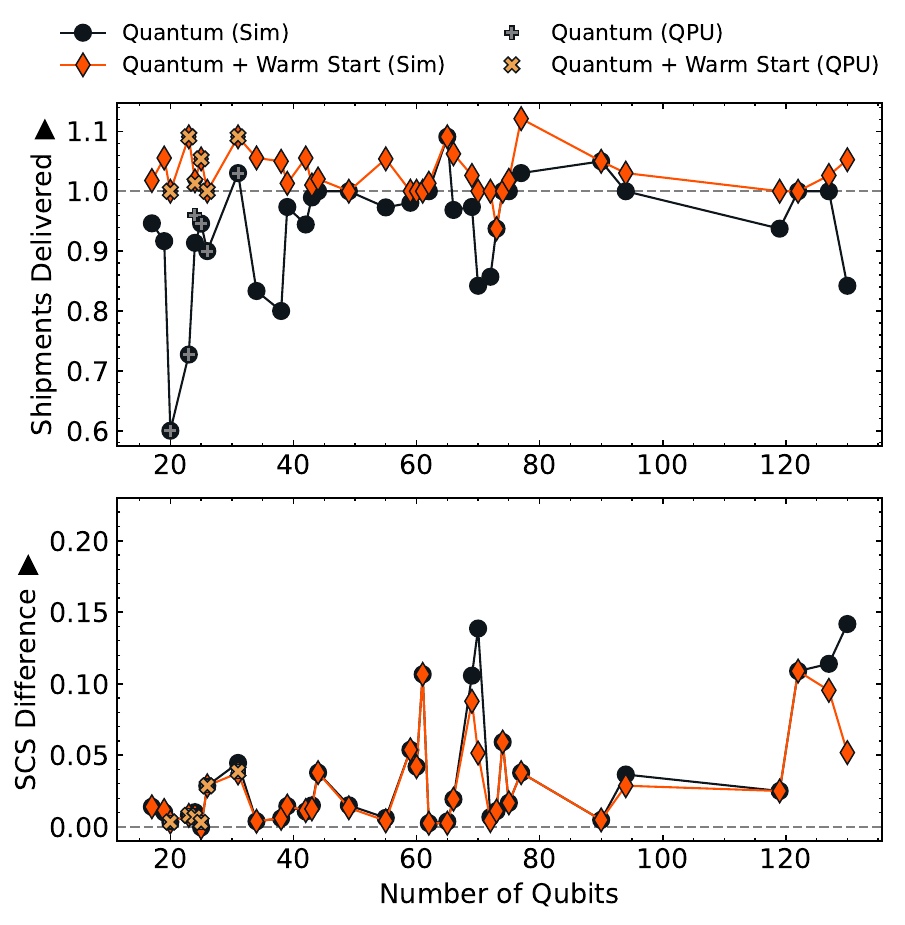}
    \caption{Best scenario-level KPI improvement versus instance size (number of qubits). For each cancellation scenario, the best reconstructed quantum or quantum warm-start solution is compared against the corresponding baseline EOS solution for that same scenario. The SD panel reports the ratio $\mathrm{SD}/\mathrm{SD}_{\mathrm{BL}}$, while the SCS panel reports the difference $\mathrm{SCS}-\mathrm{SCS}_{\mathrm{BL}}$. Markers labeled ``(QPU)'' show the same quantities for the 20--35 qubit scenarios executed on quantum hardware.}

    \label{fig:kpi_sd_max-pr_nq}
\end{figure}

\textbf{Shipments Delivered.}
The number of delivered shipments is indirectly influenced by the objective, since selecting more sequences means more shipments attempted. \cref{fig:kpi_sd} shows that the largest shipment gains arise after the quantum assignment is embedded into the hybrid workflow. The strongest QWS improvement reaches $+12.1\%$ in June, and positive best-case QWS gains appear in most monthly schedules. Even the direct reconstructed quantum solutions occasionally outperform the baseline, with best-case gains up to $+9.1\%$ in October, although their monthly averages remain negative overall. The averages therefore tell a more conservative story: across all scenarios, Q has a mean SD change of $-5.1\%$, whereas QWS recovers a positive mean SD improvement of $+1.7\%$. This gap between average and best behavior indicates that the raw quantum assignment alone does not reliably maximize delivered shipments, but it can provide high-quality starting points that EOS is able to refine into operationally stronger plans. The SD panel of \cref{fig:kpi_sd_max-pr_nq} shows that these strongest shipment gains remain available across the full range of simulated instance sizes rather than being confined to the smallest problems.

\textbf{Schedule Compatibility Score.}
The schedule compatibility score is directly captured by the quadratic term of the objective. The strongest scenario-level gains reach $+0.14$ for Q and $+0.11$ for QWS, both occurring in July, with additional best-case values of order $0.04$--$0.06$ appearing in several other months. By contrast, the monthly averages are smaller: both Q and QWS have mean $\Delta$SCS values of about $+0.02$ across the full benchmark set. This behavior is consistent with \cref{fig:kpi_scs}, where most scenario-level points are nonnegative and the largest positive deviations come from a subset of especially compatibility-favorable scenarios. The main takeaway is therefore that Iterative-QAOA reliably identifies compatibility-aware assignments, while the strongest of these gains persist after the EOS warm-start step.

\textbf{Routing efficiency.}
The routing-efficiency metrics, indirectly promoted by the quadratic term's reward for geographically clustered assignments, show the same pattern: the strongest improvements are substantially larger than the averages. For TDD, the best reconstructed quantum solutions achieve reductions as large as $-32.5\%$ in July and $-25.4\%$ in May, while the best QWS reductions reach $-9.4\%$ in June. For TDD/SD, the strongest reductions are $-6.8\%$ for Q in May and $-5.9\%$ for QWS in February. These values show that the hybrid workflow can incorporate quantum-selected assignments with much more efficient routing than the cold-start baseline. The average effects are more modest: across all scenarios, Q yields a mean TDD reduction of $-3.5\%$ but still increases TDD/SD on average by $+1.9\%$, whereas QWS has a mean TDD change of $+1.5\%$ and a nearly neutral mean TDD/SD change of $-0.2\%$. Thus, the principal routing benefit again appears in the strongest retained candidates rather than in the monthly averages alone.

\textbf{Operational cost.}
Although TC is not included in \cref{tab:kpi_avg_improvements}, we monitor it as a consistency check on the hybrid workflow. This value is indirectly minimized through the per-sequence cost-saving linear terms of the objective. Across the evaluated weekly schedules, total operational cost remains effectively unchanged relative to the baseline, indicating that the shipment and compatibility gains obtained after warm-start do not come at a material cost penalty.

\textbf{Hardware execution.}
For the 20--35 qubit scenarios executed on quantum hardware, the QPU markers in \cref{fig:kpi_sd,fig:kpi_scs,fig:kpi_sd_max-pr_nq} fall nearly on top of their simulated counterparts in both SD and SCS, indicating that the operational KPI conclusions hold under direct hardware execution using only the default debiasing setting. A dedicated instance-level comparison is provided in the Supplementary Information (\cref{sec:SI_hardware}).

\section{Conclusions and Outlook}\label{sec:future-outlook}
We have presented a hybrid quantum-classical workflow for the SSP in electric freight logistics. The SSP is formulated as a MIQP whose linear terms encode the marginal value of individual insertions and whose quadratic terms encode pairwise schedule compatibility. This formulation is mapped to a QUBO and corresponding cost Hamiltonian, solved with Iterative-QAOA, passed through an inexpensive classical in-house developed refinement, and integrated into Einride's operational EOS pipeline through reconstruction and warm-start refinement.

Our study shows that the strongest operational gains arise from the hybrid workflow built around the quantum assignment. When the quantum-derived solution is passed to EOS as a warm start, the resulting QWS plans achieve a mean SD improvement of $+1.7\%$ and a mean reduction of $0.2\%$ in drive distance per shipment across all weekly schedules and scenarios, while total operational cost remains effectively unchanged. On individual scenarios the gains are substantially larger, with best observed improvements of $+12.1\%$ in SD, $-9.4\%$ in TDD, and $-5.9\%$ in TDD/SD. These results indicate that the quantum assignment supplies a structured, compatibility-aware initialization that EOS can refine into operationally stronger plans than a cold start.

At the same time, the direct reconstructed quantum solutions remain important for understanding the source of this advantage. Before warm-start refinement, they already improve schedule compatibility by up to  $+0.14$ in SCS relative to the baseline, showing that Iterative-QAOA identifies geographically and temporally coherent insertions. Although these raw quantum solutions do not on average maximize delivered service on their own, they capture the quadratic structure that the downstream warm-start exploits.

The supplementary performance analysis further shows that Iterative-QAOA remains robust across a broad range of instance sizes without instance-by-instance parameter optimization. Across the full benchmark set, the method uses only two values of the LR-QAOA slope parameter $\Delta$, together with fixed circuit depth within each size regime, yet still attains high-quality solutions and often matches the SCIP optimum before or after refinement. This robustness is especially notable because Iterative-QAOA is non-variational and therefore avoids the outer classical optimization loop required by standard QAOA.

Several directions remain open. First, the current compatibility model uses only pairwise interactions; extending it to higher-order terms could capture richer multi-gap routing structure at the cost of a larger optimization problem and possibly much deeper quantum circuits. Second, the quadratic weight $\lambda_Q$ is set heuristically and could be calibrated more systematically against downstream operational objectives. Third, the present large-instance simulations rely on MPS approximations with fixed bond dimension, so part of the performance loss on the largest instances may reflect tensor network truncation rather than intrinsic algorithmic limitations. In that sense, larger quantum hardware with sufficiently high fidelity could be even more beneficial, since direct quantum execution would avoid the inherent approximation error of the MPS representation. We have additionally validated the workflow on IonQ trapped-ion quantum hardware for instances in the 20--35 qubit regime, where direct execution reproduces the simulation results; the corresponding instance-level comparison is provided in the Supplementary Information (\cref{sec:SI_hardware}). Extending hardware execution to the largest instances is a natural next step as device size and fidelity continue to improve. The results of this study, however, strongly indicate that near-term quantum optimization is most promising when treated not as a stand-alone replacement for classical planning, but as a structured and integrated decision module inside a larger quantum-classical hybrid optimization pipeline representing a promising path toward delivering superior performance and substantial gains for large-scale industrial logistics problems.


\section{Supplementary Information}\label{SI}

\subsection{Performance of Iterative-QAOA}
\label{sec:SI_iterqaoa_performance}

\begin{figure}[t!]
    \centering
    \includegraphics[width=\columnwidth]{
    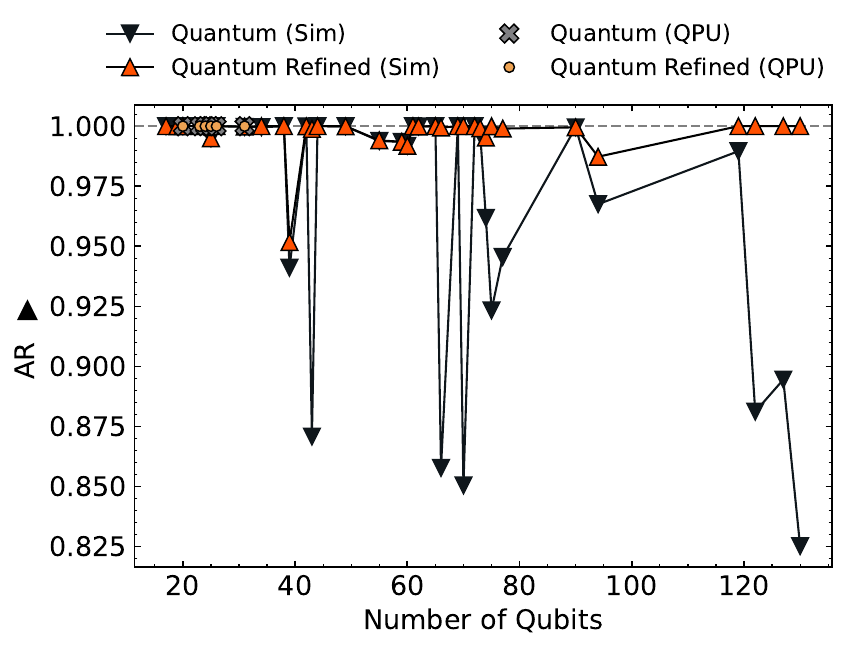}
    \caption{Approximation ratio as a function of instance size for raw Iterative-QAOA solutions and for the refined solutions obtained by applying the in-house refinement heuristic of \cref{sec:refinement}. For each instance, the approximation ratio is defined as the lowest cost Hamiltonian energy found divided by the corresponding minimum energy obtained by SCIP. Values near unity indicate agreement with the SCIP optimum. Markers labeled ``(QPU)'' show the corresponding raw and refined solutions from direct execution on IonQ trapped-ion hardware (20--35 qubits).}
    \label{fig:ar_vs_num_qubits}
\end{figure}

To summarize solution quality across the full benchmark set with a single size-dependent metric, we use the approximation ratio AR, which compares the best energy found by Iterative-QAOA against the corresponding classical optimum. AR is defined as follows
\begin{align}
    \mathrm{AR}
    =
    \frac{\min_{\mathbf{x}\in \mathcal{X}_{Q}} E(\mathbf{x})}
         {E_{\mathrm{SCIP}}},
    \label{eq:ar_definition}
\end{align}
where $\mathcal{X}_{Q}$ denotes the set of bit strings sampled by Iterative-QAOA across all iterations, $E(\mathbf{x})=\mel{\mathbf{x}}{H_C}{\mathbf{x}}$ is the corresponding cost Hamiltonian energy, and $E_{\mathrm{SCIP}}$ is the minimum energy obtained by the SCIP solver for the same instance. The behavior of \cref{fig:ar_vs_num_qubits} is consistent with the expectation that the approximation ratio of the raw Iterative-QAOA solutions should gradually decrease as the instances become larger. Several effects can contribute to this trend. First, for larger instances the simulations rely on an MPS approximation with fixed bond dimension $\chi=256$. Second, the QAOA ansatz itself may become insufficiently expressive at fixed circuit depth, so that deeper circuits could in principle reach lower energies on some of the larger instances. The decline is therefore not unexpected, and it should be interpreted as the combined effect of ansatz limitations and tensor network truncation.

At the same time, the figure highlights a useful robustness property of the approach. Across the full benchmark set, the simulations use only two values of the LR-QAOA slope parameter $\Delta$, selected solely according to instance size as described in \cref{sec:application}, together with a fixed bond dimension and a fixed choice of circuit depth within each size regime. Despite this deliberately simple parameter rule, Iterative-QAOA still reaches very high-quality solutions and, for many instances, matches the SCIP optimum exactly. In that sense, the method does not appear to rely on fine-tuned, instance-by-instance parameter optimization in order to remain competitive.

The refined solutions data in \cref{fig:ar_vs_num_qubits} further show that the residual gap in the raw quantum solutions can often be closed by a lightweight classical post-processing step. Applying the refinement heuristic of \cref{sec:refinement} to the sampled Iterative-QAOA outputs raises the approximation ratio to unity for most instances and leaves only small deviations in the remaining cases. Since this refinement stage is computationally inexpensive compared with solving the full MIQP from scratch, it provides a natural hybrid workflow in which the quantum routine supplies a high-quality starting point and the classical heuristic performs a local cleanup.

Taken together, these results provide evidence that Iterative-QAOA scales reasonably well over the instance sizes considered here while avoiding the outer variational optimization loop required by standard QAOA. The data do not establish a formal scaling law, but they do indicate that a fixed-schedule, non-variational Iterative-QAOA workflow can continue to produce useful solutions as the problem size increases, especially when paired with a modest classical refinement step.

\subsection{Classical Computation Cost}

To characterize the classical computational cost of the SSP, we performed an additional runtime study with SCIP on scenario families derived from the same eight weekly schedules used in the main text. For this analysis, we varied the quadratic weight $\lambda_Q$ and, solely for the purpose of probing classical scaling, also generated larger scenario instances from the same weekly schedules. All runs used SCIP in its standard configuration with a 48-hour wall-clock time limit.

\begin{figure}[t!]
    \centering
    \includegraphics[width=\columnwidth]{
    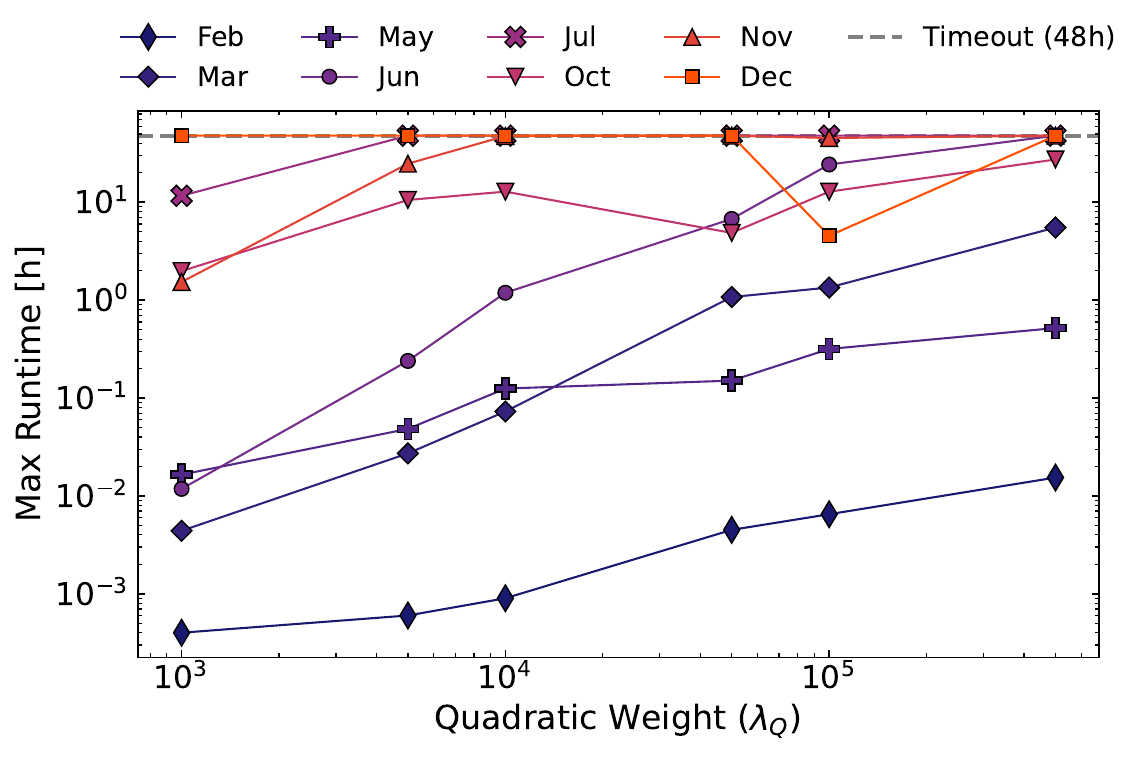}
    \caption{Maximum SCIP runtime as a function of quadratic weight $\lambda_Q$ (log--log scale), with one curve per weekly schedule. For each month label, the plotted value is the worst-case runtime across the associated scenario family derived from that weekly schedule, including the larger scenarios generated for the runtime study. The dashed gray line marks the 48-hour wall-clock time limit. Runtime varies substantially across schedules: some remain below one hour across the full parameter range, whereas others approach the timeout limit for most values of $\lambda_Q$.}
    \label{fig:classical_runtime_vs_qw}
\end{figure}

\begin{figure*}[t!]
    \centering
    \includegraphics[width=\textwidth]{
    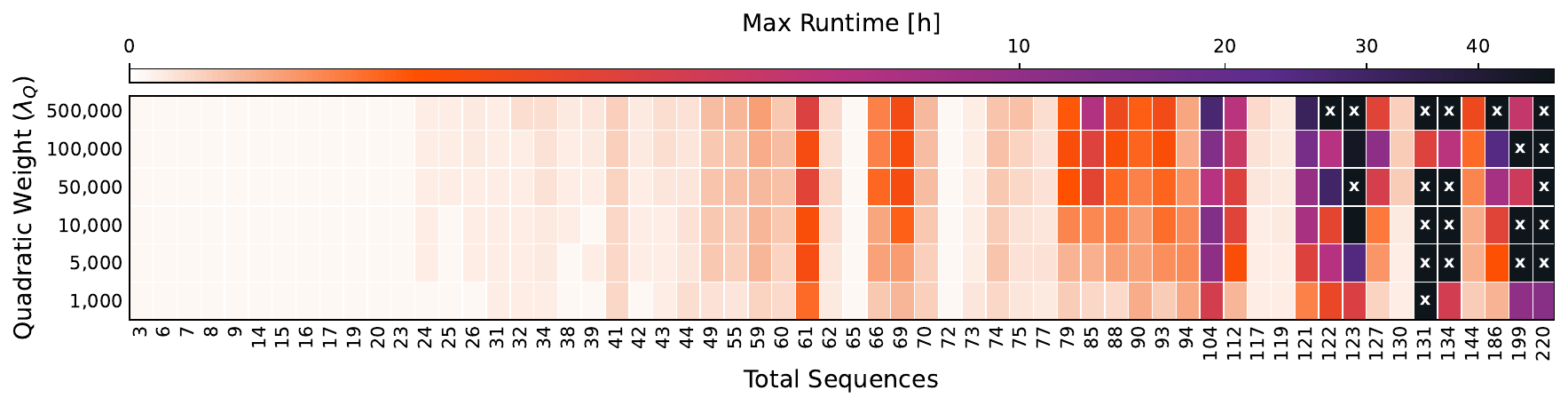}
    \caption{Heatmap of SCIP runtime over the scenario set used for the classical scaling study, with total candidate sequences $n$ on the x-axis and quadratic weight $\lambda_Q$ on the y-axis. Color encodes runtime from white (near-instant) through orange to near-black (approaching timeout), and white ``x'' markers denote runs that reached the 48-hour wall-clock limit. All scenarios are derived from the same eight weekly schedules as the main benchmark set, with additional larger scenarios generated only for runtime exploration. Runtime increases rapidly with instance size and becomes increasingly sensitive to $\lambda_Q$, with the largest scenarios frequently approaching the timeout limit.}
    \label{fig:classical_runtime_heatmap}
\end{figure*}

\cref{fig:classical_runtime_vs_qw} reports, for each monthly schedule, the worst-case runtime observed across its associated scenarios as a function of $\lambda_Q$. The figure shows that classical difficulty depends strongly on both the underlying weekly schedule and the quadratic coupling strength: some schedules remain tractable over the full parameter range, whereas others approach the timeout limit for most values of $\lambda_Q$. \cref{fig:classical_runtime_heatmap} complements this view by organizing runtimes over total candidate sequences $n$ and quadratic weight $\lambda_Q$. Taken together, the two figures indicate that classical branch-and-bound becomes increasingly expensive as the instances become larger and the quadratic structure becomes more pronounced.

\subsection{In-House Refinement Heuristic}\label{sec:SI_in_house_refinement}

\cref{alg:inhouse_refinement} gives the detailed implementation of the refinement heuristic introduced in \cref{sec:refinement}. Starting from an initial feasible assignment $S_0$, the algorithm maintains an incumbent best solution $S^\star$ and constructs a local-search trajectory by repeatedly applying feasible single-sequence \texttt{add}/\texttt{remove} moves to a working solution $S$. Candidate actions are scored by their objective change $\Delta(a)$ and are accepted only if all enforced constraints remain satisfied. During each pass, the sequence of visited feasible solutions is stored, and the incumbent is updated with the best solution observed along the trajectory rather than only with the terminal state.

Two implementation details are particularly important. First, once a sequence has been modified during a given trajectory, no further action on that same sequence is allowed in the current pass; this tabu-style restriction prevents immediate undo moves and short cycles. Second, the quadratic interaction contribution is updated incrementally through the compatibility matrix, which keeps the refinement stage inexpensive relative to solving the full MIQP from scratch. When additional iteration budget is available, the algorithm may also randomize among the top-ranked feasible actions, providing a limited form of diversification in the spirit of simple exploration strategies such as $\varepsilon$-greedy selection \cite{hajek1988cooling,geman1984stochastic,sutton1998reinforcement}.

\begin{algorithm}[tbp]
    \caption{In-House Iterative Refinement Heuristic}
    \label{alg:inhouse_refinement}
    \begin{algorithmic}[1]
    \State \textbf{Input:} instance $\mathcal{P}$, initial feasible solution $S_0$, refinement iterations $T$
    \State \textbf{Output:} best refined solution $S^\star$
    
    \State $S^\star \gets S_0$;\quad $\text{score}^\star \gets \textsc{Evaluate}(S^\star)$
    
    \For{$t=1$ to $T$}
        \State $S \gets S^\star$
        \State Initialize action table $\mathcal{A}$ over all sequences:
        \State \hspace{1em} action is \texttt{remove} if sequence is in $S$, else \texttt{add}
        \State Mark all actions as valid/invalid and compute their move values $\Delta(\cdot)$ under constraints
        \State $\mathcal{H} \gets \{(\textsc{Evaluate}(S), S)\}$
    
        \While{there exists at least one valid action in $\mathcal{A}$}
            \State $a^\star \gets \arg\max \{ \Delta(a) : a \in \mathcal{A},\ a \text{ is valid} \}$
            \State Optionally, replace $a^\star$ by a uniform choice among the top-$2$ valid actions ranked by $\Delta(a)$.
            \State Apply $a^\star$ to $S$ (update assignment, used shipments, and gap loads)
            \State Remove the sequence of $a^\star$ from $\mathcal{A}$ \Comment{tabu: each sequence moved at most once per trajectory}
            \State Recompute validity and $\Delta(\cdot)$ for remaining actions
            \State Add $(\textsc{Evaluate}(S), S)$ to $\mathcal{H}$
        \EndWhile
    
        \State $(\hat{s}, \hat{S}) \gets \arg\max_{(s,S') \in \mathcal{H}} s$
        \If{$\hat{s} > \text{score}^\star$}
            \State $S^\star \gets \hat{S}$;\quad $\text{score}^\star \gets \hat{s}$
        \Else
            \State \textbf{break} \Comment{terminate if this iteration does not improve $S^\star$}
        \EndIf
    \EndFor
    
    \State \Return $S^\star$
    \end{algorithmic}
\end{algorithm}

\subsection{Hardware Validation}\label{sec:SI_hardware}

To probe hardware performance specifically in the size regime accessible to current devices, we generated a dedicated family of instances from the same operational sweeps that underlie the main benchmark. Each instance originates from a fixed truck schedule and its set of idle gaps; an offline preprocessing step enumerates all feasible candidate sequences for every gap, ranks them per gap by a value--time efficiency score, and a per-gap cap $K$ limits how many top-ranked candidates are passed to the optimizer. Because the total number of binary decision variables---and hence qubits---equals $\sum_{g} \min(K, |\mathcal{S}_g|)$, where $|\mathcal{S}_g|$ is the number of feasible sequences for gap $g$, sweeping $K$ over a fixed schedule yields instances of different qubit counts without altering the underlying operational scenario. We use this mechanism to populate the 25--35-qubit band and execute each instance on both a noiseless simulator and IonQ trapped-ion quantum hardware, in all cases using only IonQ's default debiasing setting and no further error mitigation.

\begin{figure}[t!]
    \centering
    \includegraphics[width=\columnwidth]{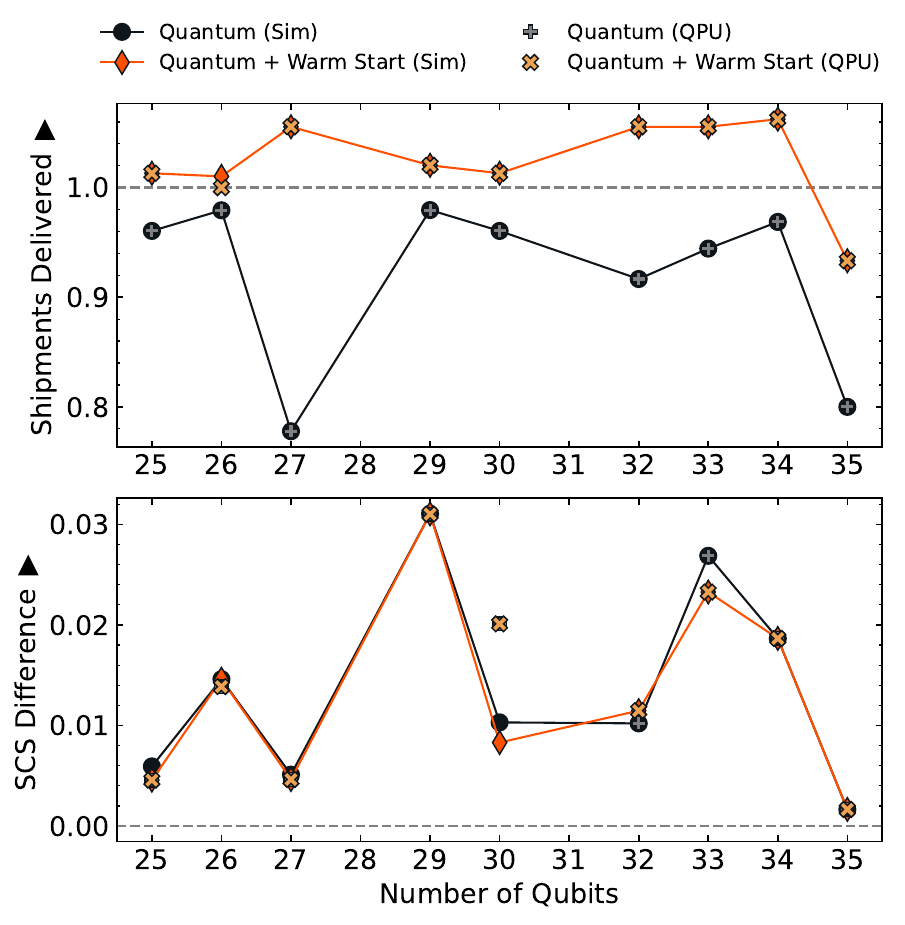}
    \caption{Hardware validation on the $K$-controlled instance family (25--35 qubits) drawn from the same operational sweeps as the main benchmark. Top: shipments delivered relative to baseline; bottom: SCS difference relative to baseline, for the reconstructed quantum (Q) and quantum warm-start (QWS) solutions. Simulation (Sim) and quantum-hardware (QPU) markers are in close agreement across all instances, frequently overlapping.}
    \label{fig:hw_validation}
\end{figure}

\Cref{fig:hw_validation} reports the resulting instance-level comparison. Across the full set, the quantum-hardware closely match the simulated values for both shipments delivered and schedule compatibility score, for the reconstructed quantum solution and for the quantum warm-start solution. The close agreement, obtained at fixed problem structure, indicates that the simulation pipeline faithfully predicts hardware behavior over this size range and that the operational conclusions of the main text are not artifacts of noiseless simulation.

\subsection{IonQ Quantum Hardware}\label{sec:SI_hardware_details}

The hardware results in this work were collected on an IonQ Forte-generation trapped-ion quantum processor \cite{Chen2024-co}. The qubit register is a chain of up to 36 $^{171}\mathrm{Yb}^{+}$ ions held in a micro-fabricated surface (Paul) trap within an integrated vacuum package, with each qubit stored in a pair of hyperfine ground-state levels of a single ion. Ions are loaded by laser ablation followed by resonant photoionization.

Gate operations are driven by counter-propagating $355\,\mathrm{nm}$ laser pulses that couple the qubit levels through two-photon Raman transitions, yielding a native gate set of arbitrary single-qubit rotations together with two-qubit $ZZ$ entangling gates. Individual-ion addressing is provided by acousto-optic deflectors that steer the control beams onto selected ions, an approach that suppresses inter-ion crosstalk and beam-pointing error \cite{Kim:2008ApOpt, Pogorelov:2021PRXQ} and is combined with automated recalibration to maintain stable operation over long experimental runs. During the measurement campaign the device operated at median two-qubit gate fidelities of approximately $99.3\%$ and single-qubit fidelities near $99.98\%$, with characteristic gate durations of about $950\,\mu\mathrm{s}$ and $130\,\mu\mathrm{s}$ for two- and single-qubit operations, respectively.

\balance
\bibliographystyle{IEEEtran}
\bibliography{bibliography}

\end{document}